\documentclass[12pt]{article}
\usepackage{amsmath}
\usepackage{amssymb}
\begin{document}
\vskip 1truein
\begin{center}
{\bf \Large{Variational principle and free falling
\\in a space-time with torsion}}
 \vskip 5pt
{Rolando Gaitan D.}${}^{\dag}$, Juan Petit and Alfredo Mej\'\i a\\
${}^\dag${\it Grupo de F\'\i sica Te\'orica, Departamento de F\'\i
sica, Facultad de Ciencias y Tecnolog\'\i a, Universidad de
Carabobo, A. P. 129
Valencia 2001, Edo. Carabobo, Venezuela.}\\
\end{center}

\vskip .2truein
\begin{abstract}

A comparison between the two possible variational principles for
the study of a free falling spinless particle in a space-time with
torsion is noted. It is well known that the autoparallel
trajectories can be obtained from a variational principle based on
a non-holonomic mapping, starting with the standard world-line
action. In a contrast, we explore a world-line action with a
modified metric, thinking about the old idea of contorsion
(torsion) potentials. A fixed-ends variational principle can
reproduce autoparallel trajectories without restrictions on
space-time torsion. As an illustration we have considered a
perturbative Weitzenb$\ddot{o}$ck space-time. The non-perturbative
problem is stablished at the end.

\end{abstract}

\vskip .2truein
\section{Introduction}

From a variational principle point of view, there exist two ways
for the study of trajectories of test particles in space-times
with curvature and torsion. A non-holonomic mapping can be
considered in the variations of a standard particle action which
provides the expected autoparallel trajectory\cite{r1}. There, the
fixed ends are not allowed and the failure of the closure of
parallelograms depends directly on the non-null torsion. Moreover,
a very particular shape of functional variations of coordinates
(i.e.,''knock variations''), must be considered in order to obtain
autoparallel trajectories.

On the other hand, for example,  the proposal of an action
constructed with the conformally transformed metric through the
0-spin part of torsion\cite{r2} is well known and, considering the
conserved Belinfante tensor and the Papapetrou idea\cite{r3}, it
is possible to obtain the autoparallel equation of motion if a
restriction on torsion is considered. This fact is noted when a
standard fixed-ends variational principle is performed in the
aforementioned type of action.

The main purpose of this work is to show that it would be possible
to extend the idea about modified metric, beyond a conformal
transformation, in the construction of a test particle action
which could reproduce the autoparallel equation of motion without
restrictions on space-time torsion. This modified metric shall
consider a conformal transformation plus auxilliary fields (i.e.,
some type of torsion potentials). The introduction of torsion
potentials is not new\cite{r4}.

This paper is organized as follows. A brief review of the
variational principle for the particle world-line action based on
a non-holonomic mapping is performed in the next section. In
section 3 we underline that there is no consistent autoparallel
equation of motion arising from the test particle action
constructed with a conformally modified metric and when a
fixed-ends variational principle takes place. This means the
appearance of serious restrictions on space-time torsion. However,
we discuss the introduction of another modified metric which
avoids any torsion restrictions and reproduce autoparallels
considering a standard variational principle. In this sense, we
tackle the most simple case as the Weitzenb$\ddot{o}$ck space-time
with a weak torsion and, an interpretation from torsion potentials
idea is performed. Suggestion on the non-perturbative formulation
is explored thinking about some type of holonomies of a connection
based on $GL(N,R)\times GL(N,R)$ for the expression of the
modified metric. We end with some remarks.

\vskip .2truein
\section{A non-holonomic map}
A brief review on a special variational principle for a spinless
test particle in a space-time with torsion\cite{r1} starts here.
Let ${e_\mu}^a$ be the vielbein which (locally) connects
curvilinear, $x^\mu$ ($\mu=0,\ldots,N-1$) and Lorentzian, $\xi^a$
($a=0,\ldots,N-1$) coordinates, in other words
\begin{equation}
{e_\mu}^a(x)=\frac{\partial \xi^a}{\partial x^\mu}\bigg|_x \,\, ,
\label{e1}
\end{equation}
and this allows us to (locally) relate the metric $g_{\mu\nu}$
with the Minkowski one, $\eta_{ab}$ as follows
\begin{equation}
g_{\mu\nu}={e_\mu}^a{e_\nu}^b \eta_{ab}\,\, . \label{e2}
\end{equation}
Indexes $\mu,\nu,\lambda,\ldots$ and $a,b,c,\ldots$ are raised and
lowered with the corresponding metrics and $\eta
=diag(-1,1,1,\ldots,1)$. The non-holonomic behavior of the map
$e:\xi^a \rightarrow x^\mu$ which is realized through the Jacobian
elements (\ref{e1}), is established by means of the fact that the
first does not satisfy the Schwarz condition (i.e., $\partial_\mu
{e_\nu}^a-\partial_\nu {e_\mu}^a\neq 0$) in presence of torsion.
So, from now on we asume an non-Riemannian space-time with null
metricity,  this means
\begin{equation}
D_\mu {e_\nu}^a=\partial_\mu {e_\nu}^a- {(A_\mu)^\lambda}_\nu
{e_\lambda}^a +{(\omega_\mu)^a}_b {e_\nu}^b=0\,\, , \label{e3}
\end{equation}
where $A_\mu$ and $\omega_\mu$ are the affine and spin
connections, respectively. Using (\ref{e3}), one can write down
the Schwarz condition as follows
\begin{equation}
{e^\rho}_a[\partial_\mu {e_\nu}^a-\partial_\nu {e_\mu}^a]=
{T^\rho}_{\mu\nu}- {t^\rho}_{\mu\nu} \neq 0\,\, , \label{e4}
\end{equation}
where ${T^\rho}_{\mu\nu}\equiv {(A_\mu)^\rho}_\nu-{(A_\nu)^\rho
}_\mu$ and ${t^\rho}_{\mu\nu}\equiv
{(\omega_\mu)^\rho}_\nu-{(\omega_\nu)^\rho }_\mu$ are the
components of the torsion tensors (affine and spin, respectively).
The affine connection is related with the Christoffel symbols
through
\begin{equation}
{(A_\mu)^\lambda}_\nu =
{\Gamma^\lambda}_{\mu\nu}(g)+{K^\lambda}_{\mu\nu}\,\, , \label{e5}
\end{equation}
defining the Christoffel symbols as always,
${\Gamma^\lambda}_{\mu\nu}(g)=\frac{g^{\rho\lambda}}{2}[\partial_\mu
g_{\rho\nu} +\partial_\nu g_{\rho\mu} -\partial_\rho g_{\mu\nu}]$
and  ${K^\lambda}_{\mu\nu}=\frac{1}{2}({T^\lambda}_{\mu\nu}
+{{T_\mu}^\lambda}_\nu+{{T_\nu}^\lambda}_\mu)$ are the contorsion
tensor components.

Let $\tau$ be an invariant parameter which allows us to describe
the particle trajectory. Then, one can write down the integral
form of the inverse transformation espressed in (\ref{e1}), in the
way
\begin{equation}
x^\mu(\tau)= x^\mu(\tau_0)+ \int_{\tau_0}^\tau
d{\tau}'{e^\mu}_a({\tau}')\dot{\xi}^a({\tau}')\,\, . \label{e6}
\end{equation}
If arbitrary functional variations on coordinates are performed
(i.e, $x'^\mu=x^\mu+\delta x^\mu$ and $\xi'^a=\xi^a+\delta\xi^a$),
from (\ref{e6}) we can write $\delta x^\mu(\tau)=
{e^\mu}_a({\tau})\delta\xi^a({\tau})-{e^\mu}_a({\tau}_0)\delta\xi^a({\tau}_0)
+\int_{\tau_0}^\tau
d{\tau}'\partial_\alpha{e^\mu}_a[\dot{\xi}^a\delta x^\alpha
-\dot{x}^\alpha \delta\xi^a]$. Thinking in a fixed starting point
(i.e., $\delta\xi^a({\tau}_0)=0$), we have
\begin{equation}
\delta x^\mu(\tau)= {e^\mu}_a({\tau})\delta\xi^a({\tau})
+\int_{\tau_0}^\tau d{\tau}'[{e_\alpha}^a\partial_\nu{e^\mu}_a
-{e_\nu}^a\partial_\alpha{e^\mu}_a]\dot{x}^\alpha \delta x^\nu\,\,
. \label{e7}
\end{equation}
Here, ${e^\mu}_a\delta\xi^a$ is the holonomic variation of
$x^\mu$, whereas $\delta x^\mu$ is the non-holonomic one.
According to  (\ref{e7}) it is confirmed the non-holonomic aspect
of the map $e:\xi^a \rightarrow x^\mu$ because this fact depends
on the contribution of the difference of affine and spin torsions,
on virtue of (\ref{e4}). In the next discussion we shall assume a
null spin torsion (i.e., ${t^\rho}_{\mu\nu}=0$).

Now, we review the world-line action for a test particle
\begin{equation}
S= \frac{m}{2}\int_{\tau_0}^\tau
d{\tau}'g_{\mu\nu}(x(\tau'))\dot{x}^\mu(\tau')
\dot{x}^\nu(\tau')\,\, , \label{e8}
\end{equation}
and considering the ''knock'' variation,
${e^\mu}_a(\tau)\delta\xi^a(\tau)=\epsilon^\mu
(\tau)\delta(\tau-\tau_0)$, the variation of the action (\ref{e8})
is $\delta S= -m\epsilon_\mu\big[\ddot{x}^\mu
+{(A_\alpha)^\mu}_\beta \dot{x}^\alpha\dot{x}^\beta\big]$. For
arbitray functions $\epsilon_\mu (\tau)$, the extremal of the
action conduce to the well known autoparallel equation
\begin{equation}
\ddot{x}^\mu +{(A_\alpha)^\mu}_\beta
\dot{x}^\alpha\dot{x}^\beta=0\,\, . \label{e9}
\end{equation}
Then, one can question if it is viable to construct a test
particle action which can reproduce  (\ref{e9}) under a fixed-ends
variational principle. This is the main purpose of the next
section.

\vskip .2truein
\section{A modified metric}
It is well known the way to obtain autoparallel equations of
motion through the Papapetrou metod\cite{r3}. However, there exist
some conditions which constraint the background (i.e., 0-spin
component of torsion is not arbitrary). This aspect can be
stressed from the point of view of a fixed-ends variational
principle of the action\cite{r2}
\begin{equation}
S= \frac{m}{2}\int_{\tau_0}^\tau
d{\tau}'e^{-\phi(x(\tau'))}g_{\mu\nu}(x(\tau'))\dot{x}^\mu(\tau')
\dot{x}^\nu(\tau')\,\, , \label{e10}
\end{equation}
where $\phi(x)$ would be thought as an arbitrary scalar field.
Notation is introduced for the conformally modified metric,
$\overline{g}_{\mu\nu}(x)=e^{-\phi(x)}g_{\mu\nu}(x)$, then the
extremal of  (\ref{e10}) provide the following equation of motion
\begin{equation}
\ddot{x}^\mu +{\Gamma^\mu}_{\alpha\beta}(\overline{g})
\dot{x}^\alpha\dot{x}^\beta=0\,\, , \label{e11}
\end{equation}
with
\begin{equation}
{\Gamma^\mu}_{\alpha\beta}(\overline{g})={\Gamma^\mu}_{\alpha\beta}(g)
-\frac{1}{2}({\delta^\mu}_\alpha
\partial_\beta\phi +{\delta^\mu}_\beta
\partial_\alpha\phi- g_{\alpha\beta}\partial^\mu\phi)\,\, . \label{e11a}
\end{equation}
If one claim that (\ref{e11}) describe autoparallel trajectories,
then the following $N$ constraints on contorsion (torsion)
components must arise
\begin{equation}
N{K_{\mu\alpha}}^{\alpha}+(N-2){K^\alpha}_{\{\mu\alpha\}}=0 \,\, ,
\label{e12}
\end{equation}
where $N$ is the dimension of the space-time and
${K^\lambda}_{\{\alpha\beta\}}\equiv\frac{1}{2}({K^\lambda}
_{\alpha\beta}+{K^\lambda}_{\beta\alpha})$. As an illustration,
let us consider a $2+1$ dimensional space-time where, if
$K_{\mu\nu}=K_{\nu\mu}$ is a symmetric tensor and
$\varepsilon^{\alpha\beta\lambda}$ is the Levi-Civita tensor; then
the contorsion tensor can be decomposed as
${K^\lambda}_{\alpha\beta}={\varepsilon^{\sigma\lambda}}_\nu
K_{\mu\sigma}+ {\delta^\lambda}_\mu V_\nu- g_{\mu\nu}V^\lambda$.
Using the last expression and (\ref{e11a}) in (\ref{e12}) conduce
to $V_\alpha=0$ and $\phi=0$, which is a serious restriction on
the model.

Next, we explore an extension of the modified metric definition
and the action is given by
\begin{equation}
S= \frac{m}{2}\int_{\tau_0}^\tau
d{\tau}'\widetilde{g}_{\mu\nu}(x(\tau'))\dot{x}^\mu(\tau')
\dot{x}^\nu(\tau')\,\, , \label{e13}
\end{equation}
where notation means
\begin{equation}
\widetilde{g}_{\mu\nu}(x)=e^{-\phi(x)}g_{\mu\nu}(x)+
h_{\mu\nu}(x)+ {\mathcal{O}^2}_{\mu\nu}(\phi,h)(x) \,\, .
\label{e14}
\end{equation}
Here, $\phi(x)$ and $h_{\mu\nu}(x)$ are scalar and symmetric
rank-two field, respectively. The additional term
${\mathcal{O}^2}_{\mu\nu}(\phi,h)(x) $ represents all possible
contributions on higher order of auxiliary fields. The extremized
action (\ref{e13}) leads to
\begin{equation}
\ddot{x}^\mu +{\Gamma^\mu}_{\alpha\beta}(\widetilde{g})
\dot{x}^\alpha\dot{x}^\beta=0\,\, , \label{e15}
\end{equation}
and if we claim that this equation describes autoparallels, the
following condition appears
\begin{equation}
{K^\mu}_{\{\alpha\beta\}}=
{\Gamma^\mu}_{\alpha\beta}(\widetilde{g})-{\Gamma^\mu}_{\alpha\beta}(g)\,\,
. \label{e16}
\end{equation}
The main task is to solve (\ref{e16}) obtaining possible solutions
for auxiliary fields, $\phi(x)$ and $h_{\mu\nu}(x)$ in terms of
the semi-symmetric components of contorsion and without
constraints on the last one.

As an illustration, let us consider a Weitzenb$\ddot{o}$ck
space-time with a weak torsion and the natural relationship,
${K^\lambda}_{\mu\nu}\rightarrow 0 \Rightarrow \phi\rightarrow 0 ,
h_{\mu\nu}\rightarrow 0$. At first order we write
\begin{equation}
\widetilde{g}_{\mu\nu}=\eta_{\mu\nu}
-\phi\eta_{\mu\nu}+h_{\mu\nu}\,\, , \label{e17}
\end{equation}
\begin{equation}
\widetilde{g}^{\mu\nu}=\eta^{\mu\nu} +\phi\eta^{\mu\nu}-h^{\mu\nu}
\,\, , \label{e18}
\end{equation}
where the indexes are lowered or raised with the Minkowski metric
and
$\widetilde{g}_{\mu\nu}\widetilde{g}^{\mu\lambda}={\delta_\nu}^\lambda$.
With the help of (\ref{e17}) and (\ref{e18}), we can rewrite
(\ref{e16}) as follows
\begin{equation}
{K^\mu}_{\{\alpha\beta\}}=
{\gamma^\mu}_{\alpha\beta}(h)-\frac{1}{2}({\delta^\mu}_\alpha
\partial_\beta\phi +{\delta^\mu}_\beta
\partial_\alpha\phi- \eta_{\alpha\beta}\partial^\mu\phi)\,\, , \label{e19}
\end{equation}
where  ${\gamma^\mu}_{\alpha\beta}(h)\equiv
\frac{\eta^{\mu\sigma}}{2}[\partial_\alpha h_{\sigma\beta}
+\partial_\beta h_{\sigma\alpha} -\partial_\sigma
h_{\alpha\beta}]$. A trace of (\ref{e19}) gives us the first
derivative of scalar field $\phi$, in other words
\begin{equation}
\partial_ \beta \phi= \frac{2}{N}({K^\mu}_{\{\mu\beta\}}-
{\gamma^\mu}_{\mu\beta}(h))\,\, , \label{e20}
\end{equation}
and (\ref{e19}) is now
\begin{eqnarray}
\gamma_{\mu\alpha\beta}(h)-\frac{1}{N}\big(\eta_{\mu\beta}
{\gamma^\sigma}_{\sigma\alpha}(h)+\eta_{\mu\alpha}{\gamma^\sigma}_{\sigma\beta}(h)-
\eta_{\alpha\beta}{\gamma^\sigma}_{\sigma\mu}(h)\big)\nonumber \\
=
K_{\mu\{\alpha\beta\}}-\frac{1}{N}\big(\eta_{\mu\beta}{K^\sigma}
_{\{\sigma\alpha\}}+\eta_{\mu\alpha}{K^\sigma}_{\{\sigma\beta\}}-
\eta_{\alpha\beta}{K^\sigma}_{\{\sigma\mu\}}\big)\,\, .\label{e21}
\end{eqnarray}
It can be observed that the right-hand side of (\ref{e21}) is the
traceless part of the semi-symmetric contorsion; this means
${K^t}_{\mu\{\alpha\beta\}}\equiv
K_{\mu\{\alpha\beta\}}-\frac{1}{N}\big(\eta_{\mu\beta}{K^\sigma}
_{\{\sigma\alpha\}}+\eta_{\mu\alpha}{K^\sigma}_{\{\sigma\beta\}}-
\eta_{\alpha\beta}{K^\sigma}_{\{\sigma\mu\}}\big)$ with
${K^{t\mu}}_{\{\mu\beta\}}\equiv 0$. So, if we write the traceless
part of the field $h_{\mu\nu}$ in the way
$h_{\alpha\beta}={h^t}_{\alpha\beta}
+\frac{\eta_{\alpha\beta}}{N}\,{h^\sigma}_\sigma$ with
${h^{t\sigma}}_{\sigma}\equiv 0$, equation (\ref{e21}) is
\begin{equation}
\gamma_{\mu\alpha\beta}(h^t) = {K^t}_{\mu\{\alpha\beta\}}\,\,
,\label{e23}
\end{equation}
saying that the trace of ${h^\sigma}_\sigma $ remain unsolved. The
solution for ${h^t}_{\alpha\beta}$ from  (\ref{e23}) is a
straightforward task, in other words
\begin{equation}
\partial_\mu{h^t}_{\alpha\beta} = {K^t}_{\alpha\{\beta\mu\}}+{K^t}_{\beta\{\alpha\mu\}} \,\,
.\label{e24}
\end{equation}

Up to a fixation of the trace of $h_{\mu\nu}$, it is possible to
obtain all the auxiliary fields using (\ref{e20}) and (\ref{e24}).
In a linearized regime, it would be sufficient to choose
$\widetilde{g}_{\mu\nu}=\eta_{\mu\nu}
-\phi\eta_{\mu\nu}+{h^t}_{\mu\nu}$ instead of (\ref{e17}).

In the non-perturbative case, equation (\ref{e16}) represents a
homogeneous first order diferential equations system and, assuming
a non singular metric $\widetilde{g}_{\mu\nu}$, one can write
\begin{equation}
\partial_\beta\widetilde{g}_{\lambda\alpha} -\big[
{\delta^\sigma}_\lambda{(A_{\{\alpha})^\mu}_{\beta\}}
+{\delta^\sigma}_\alpha{(A_{\{\lambda})^\mu}_{\beta\}}\big]\widetilde{g}_{\sigma\mu}=0
\,\, .\label{e25}
\end{equation}
This equation suggests solutions for $\widetilde{g}_{\mu\nu}$ via
some type of holonomies of some connection and, then an algebraic
problem for the solutions of auxiliary fields (potentials)  $\phi$
and $h_{\mu\nu}$, considering (\ref{e14}). This business related
to holonomies can be focused from the following point of view. If
one considers  composed indexes (i.e., Petrov indexes), $A$, $B$,
$C$,...$=$ $\mu \nu$ and the elements of the transformation group
for greek ones are $U\in GL(N,R)$, then, if the transformation
rule for the metric is $\widetilde{g}'_{\mu\nu}={U^\alpha}_\mu
{U^\beta}_\nu \widetilde{g}_{\alpha\beta}$, it can be written as
follows
\begin{equation}
\widetilde{g}'_A={\mathcal{U}^B}_A  \widetilde{g}_B \,\,
,\label{e26a}
\end{equation}
where ${\mathcal{U}^B}_A \equiv {U^{\alpha}}_{\mu}
{U^{\beta}}_{\nu}$ $\in$ $\mathcal{G}\equiv GL(N,R)\times GL(N,R)$
and the metric $\widetilde{g}_B$ is arranged like a vector with
$N^2$ components. Next, let
${(\mathcal{A}_{\beta})^{\sigma\mu}}_{\lambda\alpha}\equiv
{(\mathcal{A}_{\beta})^A}_B$ be a field defined by
\begin{equation}
{(\mathcal{A}_\beta)^{\sigma\mu}}_{\lambda\alpha}\equiv
{\delta^{\{\sigma}}_\lambda
{(A_{\{\alpha})^{\mu\}}}_{\beta\}}+{\delta^{\{\sigma}}_\alpha
{(A_{\{\lambda})^{\mu\}}}_{\beta\}}  \,\, ,\label{e26}
\end{equation}
equation (\ref{e25}) is now
\begin{equation}
\partial_\mu\widetilde{g}_A
-{(\mathcal{A}_{\mu})^B}_A \widetilde{g}_B=0\,\, ,\label{e26b}
\end{equation}
and covariance of this equation is satisfied if
$\mathcal{A}_{\mu}$ transforms as a connection under
$\mathcal{G}$, this means
\begin{equation}
{(\mathcal{A}_\mu)^{B'}}_{A'}={(\mathcal{U}^{-1})^B}_C{(\mathcal{A}_\mu)^C}_D{\mathcal{U}^D}_A+
{(\mathcal{U}^{-1})^B}_C\partial_\mu{\mathcal{U}^C}_A \,\,
.\label{e26c}
\end{equation}

Let ${P_{x_o}}^x$ be an oriented path which starts at ''$x_o$''
and ends at ''$x$'', the solution of (\ref{e26b}) is
\begin{equation}
\widetilde{\mathbf{g}}(x)=
\mathrm{P}\,exp\big(\int_{{P_{x_o}}^x}dx^\beta
\mathcal{A}_{\beta}\big)\,\,\widetilde{\mathbf{g}}(x_o) \,\,
,\label{e27}
\end{equation}
where $\widetilde{\mathbf{g}}$ is the $N^2$-vector  with
components $\widetilde{g}_A$ and the symbol ''$\mathrm{P}$'' means
path ordered integration. Then, solutions for
$\widetilde{g}_{\alpha\beta}$ are coming from holonomies with
algebraic constraints (i. e.,
${(\mathcal{A}_\beta)^{\sigma\mu}}_{\lambda\alpha}={(\mathcal{A}_\beta)^{\mu\sigma}}_{\lambda\alpha}
={(\mathcal{A}_\beta)^{\sigma\mu}}_{\alpha\lambda}$).

\vskip .2truein
\section{Conclusion}

A possible action for a massive spinless particle when its free
falling does take place in a space-time with torsion, has been
introduced. Considering the old idea of contorsion (torsion)
potentials, the interaction is provided through a modified metric
and, by means of a standard variational principle the autoparallel
equation of motion can be recovered without restrictions on
contorsion (torsion). This fact is illustrated taking into account
the simplest and not trivial case as it is the
Weitzenb$\ddot{o}$ck space-time with a weak torsion (contorsion).
There, the potential character of auxilliary fields $\phi$ and
$h_{\mu\nu}$ is revealed when they are solved from direct
integration of semi-symmetric contorsion components. At a
non-perturbative regime, the situation is not different and the
modified metric can be obtained from a certain class of holonomies
of the connection $\mathcal{A}_{\mu}$, defined due to some
algebraic constraints. The task related to particles with spin in
a free falling in space-times with torsion must be considered
elsewhere.

\vskip .2truein




\end{document}